\documentclass[prd,letterpaper,preprint,amsmath,amsfonts,amssymb,nofootinbib,showpacs]{revtex4}
\usepackage{epsfig,psfrag,graphicx,bm}
\usepackage{dcolumn}
\begin{document}
\title{Two photon annihilation of Kaluza-Klein dark matter}
\author{Lars Bergstr\"om}
\email{lbe@physto.se}
\author{Torsten Bringmann}
\email{troms@physto.se}
\author{Martin Eriksson} 
\email{mate@physto.se} 
\author{Michael Gustafsson}
\email{michael@physto.se}
\affiliation{Department of Physics,  Stockholm University, AlbaNova
University Center, SE - 106 91 Stockholm, Sweden}
\date{\today}
\begin{abstract}
We investigate the fermionic one-loop cross section for the two photon
annihilation of Kaluza-Klein (KK) dark matter particles in a model of
universal extra dimensions (UED). This process gives a nearly
mono-energetic gamma-ray line with energy equal to the KK dark matter
particle mass. We find that the cross section is large enough that if
a continuum signature is detected, the energy distribution of
gamma-rays should end at the particle mass with a peak that is visible
for an energy resolution of the detector at the percent level. This
would give an unmistakable signature of a dark matter origin of the
gamma-rays, and a unique determination of the dark matter particle
mass, which in the case studied should be around 800 GeV.  Unlike the
situation for supersymmetric models where the two-gamma peak may or
may not be visible depending on parameters, this feature seems to be
quite robust in UED models, and should be similar in other models
where annihilation into fermions is not helicity suppressed. The
observability of the signal still depends on largely unknown
astrophysical parameters related to the structure of the dark matter
halo. If the dark matter near the galactic center is adiabatically
contracted by the central star cluster, or if the dark matter halo has
substructure surviving tidal effects, prospects for detection look
promising.
\end{abstract}
\maketitle

\newcommand{\be}{\begin{equation}}
\newcommand{\ee}{\end{equation}}
\newcommand{\bea}{\begin{eqnarray}}
\newcommand{\eea}{\end{eqnarray}}
\newcommand{\B}{B^{(1)}}

\section{Introduction}
\label{sec:intro}
The mystery of the dark matter, first indicated by Zwicky some 70
years ago \cite{zwicky}, is still with us. Thanks to remarkably
accurate measurements of the microwave background \cite{cmb},
supernova luminosity distances \cite{sn} and the redshift distribution
of baryonic structure \cite{redshift}, we now have, however, a good
handle on the required dark matter density of the Universe.

Among the various possibilities for the dark matter constituents,
weakly interacting massive particles (WIMPs) are particularly
attractive, as gauge couplings and masses in the electroweak symmetry
breaking range (50 - 1000 GeV, say) give a thermal relic abundance of
particles that naturally comes within the observed
range. Supersymmetry or Kaluza-Klein (KK) particles in models with
large extra dimensions are examples which span some of the range of
the phenomenology to expect \cite{reviews}.

Dark matter particles can in principle be detected in three different
ways: In accelerators, if the energy is high enough to produce them.
Through direct detection in ultra-sensitive detectors, where particles
from the local neighborhood of the halo deposit enough energy to be
detected.  Finally, indirectly through the contribution to the cosmic
ray flux (positrons, antiprotons, neutrinos and photons of all
energies up to the particle mass) caused by annihilations in the halo
(or in celestial bodies for neutrinos).
Here we will focus on indirect detection through high energy
gamma-rays, a signal that has recently become much more interesting
because of new large-area air Cherenkov telescopes (ACTs)
\cite{cherenkov,hess} and the launch of the Integral satellite
\cite{integral}.  In a couple of years, a new satellite, GLAST
\cite{glast}, will take the field of high energy gamma-ray
astrophysics to yet another level of accuracy and precision.

There are many reported indications of cosmic ray signals which could
in principle at least partly be due to the annihilation products of
dark matter particles. Examples are possible excesses in microwave
radiation \cite{finkenbeiner}, in positrons \cite{posexp,positrons}
and gamma-rays from the galactic center (GC) in the MeV range
\cite{mevexp,mev}, GeV range \cite{gevexp,gev,BUB} and TeV range
\cite{tevexp,tev}. Indeed, there have even been suggestions that the
combined data on positrons, gamma-rays and antiprotons could be better
explained by adding a neutralino annihilation contribution
\cite{deboer}.

The problem with these indications is that they lack distinctive
features, which will make the alternative hypothesis that there are
unknown background contributions difficult to reject. A spectacular
signature would on the other hand be the mono-energetic gamma-rays
expected from WIMP annihilation into $\gamma\gamma$ \cite{gg} or
$Z\gamma$ \cite{zg}. The phenomenology of this process has been worked
out in quite some detail for the supersymmetric case
\cite{BUB,susygg}.  Here we present the first analytical and numerical
calculations of the dominant fermion loop contributions to the process
$\B\B\rightarrow\gamma\gamma$ for KK dark matter. This dark matter
particle has recently been shown to contain a useful experimental
signature in the form of a gamma-ray spectrum that remains remarkably
flat up to the KK mass, and then steeply drops at that energy
\cite{bbeg}. Here, we investigate whether there could also be a sharp
peak at $E_\gamma=M_{\B}$ due to the two-gamma process. Needless to
say, there is no known astrophysical background which would cause a
similar structure, so the question is only if the particle physics
annihilation cross section and the dark matter halo density are large
enough to make the signal visible.

The paper is organized as follows. We begin by introducing the model
that gives rise to the KK dark matter candidate $\B$. In
Sections~\ref{pol} and \ref{cross} we then discuss the basic
properties of the cross section and give the analytical result for the
fermionic contribution. We also comment on a full numerical evaluation
of the process $\B\B\rightarrow\gamma\gamma$ and argue that the
fermionic part gives the main contribution. In Section~\ref{obs} we
insert the analytical results into a model of the Milky Way halo that
represents the current knowledge based on N-body simulations and
baryonic contraction near the GC. We also make some speculations on
the existence of dark matter substructure (``clumps'') and finally,
Section~\ref{conc}, contains our summary and conclusions.

\section{Universal extra dimensions}
\label{ued}

The lightest KK particle (LKP) is the first viable particle dark
matter candidate to arise from extra dimensional extensions of the
standard model (SM) of particle physics. It appears in models of
universal extra dimensions (UED) \cite{app,che,ser} (for the first
proposal of TeV sized extra dimensions, see \cite{anton}), where all
SM fields propagate in the higher dimensional bulk, and is stable due
to conserved KK parity, a remnant of KK mode number conservation. This
is analogous to supersymmetric dark matter models, where conserved
R-parity ensures the stability of, e.g., the neutralino. Contrary to
the supersymmetric case however, the unknown parameter space is quite
small and will be scanned throughout by next generation's accelerator
experiments.

We will consider the simplest, five dimensional model with one UED
compactified on an $S^1/Z_2$ orbifold of radius $R$. All SM fields are
then accompanied by a tower of increasingly massive KK states and at
tree-level, the $n$th KK mode mass is given by
\begin{equation}
  m^{(n)} = \sqrt{(n/R)^2 + m_\text{EW}^2}\,,
\end{equation}
where $m_\text{EW}$ is the corresponding zero mode mass. However, the
identification of the LKP is nontrivial because radiative corrections
to the mass spectrum of the first KK level are typically larger than
the corresponding electroweak mass shifts. A one-loop calculation
\cite{che} shows that the LKP is given by the first KK excitation of
the photon, which is well approximated by the first KK mode of the
hypercharge gauge boson $\B$ since the electroweak mixing angle for KK
modes is effectively driven to zero. The $\B$ relic density was
determined in \cite{ser}. Depending on the exact form of the mass
spectrum and the resulting coannihilation channels, the limit from the
Wilkinson Microwave Anisotropy Probe (WMAP) \cite{cmb} of
$\Omega_\text{CDM} h^2 = 0.12 \pm 0.02$ corresponds to $0.5\text{~TeV}
\lesssim m_{\B} \lesssim 1\text{~TeV}$. Here $\Omega_\text{CDM}$ is
the ratio of dark matter to critical density and $h$ is the Hubble
constant in units of $100\text{~km} \text{\,s}^{-1}
\text{\,Mpc}^{-1}$. Collider measurements of electroweak observables
give a current constraint of $R^{-1} \gtrsim 0.3\text{~TeV}$
\cite{app,colider}, whereas LHC should probe compactification radii up
to 1.5 TeV \cite{cheb}.

Although all SM fields come with a tower of KK states, only those
which are even under $Z_2$ have a zero mode. Fermions in five
dimensions are not chiral, but chiral zero modes can still be singled
out by assigning different transformation properties to the $SU(2)$ doublet and singlet,
respectively. Denoting four dimensional spacetime coordinates by
$x^\mu$ and the fifth coordinate by $y$, the fermion expansion in KK
states is then given by
\begin{eqnarray}
  \psi_d(x^\mu,y) &=&  \frac{1}{\sqrt{2\pi R}} P_L\psi^{(0)}(x^\mu) +
  \frac{1}{\sqrt{\pi R}} \sum_{n=1}^{\infty} \left[ P_L\psi_d^{(n)}(x^\mu)
  \cos{\frac{ny}{R}} + P_R\psi_d^{(n)}(x^\mu) \sin{\frac{ny}{R}}\right]\,,\nonumber \\
  \psi_s(x^\mu,y) &=&  \frac{1}{\sqrt{2\pi R}} P_R\psi^{(0)}(x^\mu) +
  \frac{1}{\sqrt{\pi R}} \sum_{n=1}^{\infty} \left[ P_R\psi_s^{(n)}(x^\mu)
  \cos{\frac{ny}{R}} + P_L\psi_s^{(n)}(x^\mu)\sin{\frac{ny}{R}} \right]\,,
\end{eqnarray}
where
\begin{equation}
  P_{L,R} = \frac{1\mp\gamma^5}{2}.
\end{equation}
Thus at a given KK level, the fermionic field content is doubled as
compared to the SM. In the following we neglect the zero mode fermion
mass, and set the radiative mass corrections to the singlet and
doublet KK excitations equal. The mass eigenstates are then given by
\begin{equation}
  \xi_d^{(n)} = \psi_d^{(n)}, \qquad \xi_s^{(n)} = -\gamma^5 \psi_s^{(n)},
\end{equation}
which contributes a relative minus sign in the Feynman rules for
$\B$-fermion vertices.

The number of scalar degrees of freedom is also doubled, since the
fifth components of the gauge bosons transform as four dimensional
scalars. These are required to be odd under $Z_2$ by gauge
invariance. The KK excitations of $Z_5$ and $W^\pm_5$ combine with
those of the three Goldstone bosons of the SM to form three physical
and three unphysical scalars, the latter giving masses to
$Z_\mu^{(n)}$ and ${W^\pm_\mu}^{(n)}$. The KK photon acquires a mass
by eating $A_5$, while the Higgs excitations remain physical.

\section{The $\B\B\rightarrow\gamma\gamma$ polarization tensor}
\label{pol}

Let us now discuss the direct annihilation of $\B$ particles into
photons. Taking all momenta as ingoing, the amplitude for this process
is given by
\begin{equation}
  \label{amp}
  \mathcal{M}=\epsilon^{\mu_1}_1(p_1)\epsilon^{\mu_2}_2(p_2)
  \epsilon^{\mu_3}_3(p_3)\epsilon^{\mu_4}_4(p_4)\,
  \mathcal{M}_{\mu_1\mu_2\mu_3\mu_4}(p_1,p_2,p_3,p_4)\,,
\end{equation}
where $p_{1,2}$ and $p_{3,4}$ denote the four-momenta of the $\B$s
and photons with polarization vectors $\epsilon_1,..,\epsilon_4$,
respectively.

\begin{figure}[t]
\includegraphics[width=\textwidth]{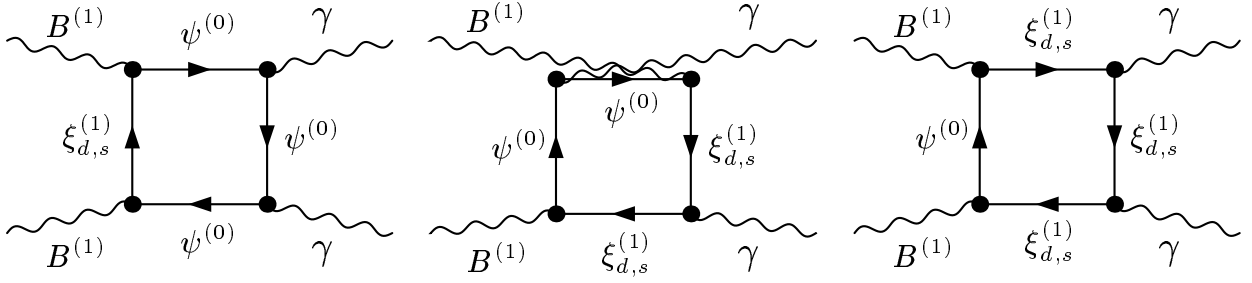}
\caption{Fermion box contributions to $\B\B \rightarrow \gamma\gamma$
including the first level of KK excitations. Not shown are the
additional nine diagrams that are obtained by crossing external
momenta.}
\label{fermions}
\end{figure}

Let us first consider those contributions to the polarization tensor
that do not contain scalars. At the one-loop level, and taking into
account only the first level of KK-excitations, they are given by the
fermion box diagrams that are shown in Fig.~\ref{fermions}. Photons
couple to the charge of the KK fermions by the usual vector coupling,
while the couplings of the $\B$s to fermions contains a vector as well
as an axial vector part and is given by
\begin{equation}
  \label{scoupling}
  -g_{Y}\frac{Y_s}{2}\B_\mu\bar \xi_s^{(1)}\gamma^\mu(1+\gamma^5)\psi^{(0)}\,+c.c.
\end{equation}
and
\begin{equation}
  \label{dcoupling}
  g_{Y}\frac{Y_d}{2}\B_\mu\bar \xi_d^{(1)}\gamma^\mu(1-\gamma^5)\psi^{(0)}\,+c.c.\,,
\end{equation}
respectively. Here, $\psi^{(0)}$ is a zero level fermion and
$\xi_s^{(1)}$ ($\xi_d^{(1)}$) its first singlet (doublet) KK
excitation with hypercharge $Y_s$ ($Y_d$) as explained above. From
charge conjugation invariance it follows that there is no total axial
vector contribution to the process (i.e. no remaining $\gamma^5$ in
the trace). This means that the polarization tensor can be written as
\begin{equation}
  \mathcal{M}^{\mu_1\mu_2\mu_3\mu_4} = -i\alpha_{em}\alpha_YQ^2(Y_s^2+Y_d^2)
  \left(\mathcal{M}_v^{\mu_1\mu_2\mu_3\mu_4}+\mathcal{M}_a^{\mu_1\mu_2\mu_3\mu_4}\right)\,,
\end{equation}
where we have pulled out a common factor for later convenience, with
$\alpha_{em}\equiv e^2/4\pi$, $\alpha_Y\equiv g_{Y}^2/4\pi$ and $Q$
being the charge of the fermions in the loop. In the above expression,
$\mathcal{M}_v$ describes the contributions from vector couplings
only, whereas $\mathcal{M}_a$ comes from those terms that only contain
$\B$ axial vector couplings. In our case, with the couplings given by
(\ref{scoupling}) and (\ref{dcoupling}),
\begin{equation}
  \mathcal{M}_v^{\mu_1\mu_2\mu_3\mu_4}=\mathcal{M}_a^{\mu_1\mu_2\mu_3\mu_4}\,.
\end{equation}

The typical velocity of WIMPs is about $v\sim10^{-3}$, i.e. they are
highly non-relativistic. The momenta of the incoming $\B$s are therefore
well approximated by $p\equiv
p_1=p_2=(m_{\B},\mathbf{0})$. Taking into account momentum
conservation,
\begin{equation}
 2\,p^\mu+p_3^\mu+p_4^\mu=0,
\end{equation}
and transversality of the polarization vectors,
\begin{equation}
  \epsilon_1\cdot p=\epsilon_2\cdot p=\epsilon_3\cdot p_3=\epsilon_4\cdot p_4=0,
\end{equation}
the Lorentz structure of the polarization tensor can be decomposed as
\begin{eqnarray}
  \mathcal{M}_v^{\mu_1\mu_2\mu_3\mu_4}
  &=&\frac{A}{m_{\B}^4}~p_3^{\mu_1}p_4^{\mu_2}p^{\mu_3}p^{\mu_4}+
   \frac{B_1}{m_{\B}^2}~g^{\mu_1\mu_2}p^{\mu_3}p^{\mu_4}+\frac{B_2}{m_{\B}^2}~g^{\mu_1\mu_3}p_4^{\mu_2}p^{\mu_4}\nonumber\\ 
  &&  +\frac{B_3}{m_{\B}^2}~g^{\mu_1\mu_4}p_4^{\mu_2}p^{\mu_3}
    +\frac{B_4}{m_{\B}^2}~g^{\mu_2\mu_3}p_3^{\mu_1}p^{\mu_4}+\frac{B_5}{m_{\B}^2}~g^{\mu_2\mu_4}p_3^{\mu_1}
   p^{\mu_3}\nonumber\\ 
  && +\frac{B_6}{m_{\B}^2}~g^{\mu_3\mu_4}p_3^{\mu_1}p_4^{\mu_2}+C_1~g^{\mu_1\mu_2}g^{\mu_3\mu_4}+C_2~g^{\mu_1\mu_3}g^{\mu_2\mu_4}+C_3~g^{\mu_1\mu_4}g^{\mu_2\mu_3}\,,
\end{eqnarray}
where $A$, $B$ and $C$ are dimensionless scalars that depend on the
external momenta and the masses appearing in the loop.

Bose symmetry prescribes that the tensor must be invariant under the
permutations $(p_1,\mu_1)\leftrightarrow(p_2,\mu_2)$ and
$(p_3,\mu_3)\leftrightarrow(p_4,\mu_4)$. This translates into
\begin{eqnarray}
  \label{crossing}
   B_2 &=& -B_4\,, \nonumber\\
   B_3 &=& -B_5\,, \nonumber\\
   C_2 &=& C_3\,, \nonumber\\
   A(3,4) &=& A(4,3)\,, \nonumber\\
   B_1(3,4) &=& B_1(4,3)\,, \nonumber\\
   B_2(3,4) &=& -B_3(4,3)\,, \nonumber\\
   B_6(3,4) &=& B_6(4,3)\,, \nonumber\\
   C_1(3,4) &=& C_1(4,3)\,, \nonumber\\
   C_2(3,4) &=& C_3(4,3)\,,
\end{eqnarray}
where we have introduced the notation $A(3,4)\equiv A(p,p_3,p_4)$,
$A(4,3)\equiv A(p,p_4,p_3)$ etc.

Because of gauge invariance the polarization tensor furthermore has to
be transversal to the photon momenta, i.e.
\begin{equation}
  \epsilon^{\mu_1}_1\epsilon^{\mu_2}_2{p^{\mu_3}_3}{\epsilon^{\mu_4}_4}\mathcal{M}_{\mu_1\mu_2\mu_3\mu_4}=\epsilon^{\mu_1}_1\epsilon^{\mu_2}_2{\epsilon^{\mu_3}_3}{p^{\mu_4}_4}\mathcal{M}_{\mu_1\mu_2\mu_3\mu_4}=0\,.
\end{equation}
This gives
\begin{eqnarray}
  \label{gauge}
  A &=& B_2-B_4-2 B_6\,,\nonumber\\
  C_1 &=& -\frac{1}{2}B_1\,,\nonumber\\
  C_2 &=& B_5\,,\nonumber\\
  C_3 &=& - B_3\,.
\end{eqnarray}

With the help of Eqs.~(\ref{crossing}) and (\ref{gauge}) one can
express the whole tensor amplitude by means of only three form
factors, which we choose to be $B_1$, $B_2$ and $B_6$. The full
analytical expressions for these functions are given in the Appendix
. We actually calculated \emph{all} the form factors $A$, $B$ and $C$
independently and used Eqs.~(\ref{crossing}) and (\ref{gauge}) merely
as a consistency check of our results.

\section{Cross section}
\label{cross}

\begin{figure}[t]
\includegraphics[width=0.7\textwidth]{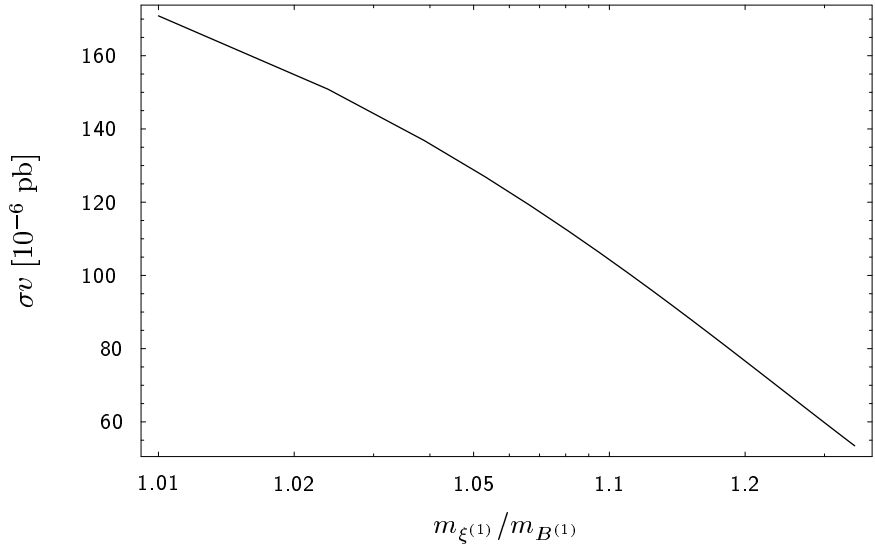}
\caption{The annihilation rate for $m_{\B}=$ 0.8 TeV as a function of
the mass shift between the $\B$ and KK fermions. The dependence on the
$\B$ mass is simply given by $\sigma v\propto m_{\B}^{-2}$.}
\label{sigmafig}
\end{figure}

For low relative velocities $v$ between the incoming $\B$ particles
the annihilation cross section $\sigma$ diverges, but the annihilation
rate is proportional to $\sigma v$ and finite. Assuming the same mass
shift for all KK fermions, we find after summing over final and
averaging over initial states:
\begin{equation}
 \label{sigmav}
 \sigma v=\frac{\alpha_Y^2\alpha_{em}^2g_{eff}^4}{144 \pi m_{\B}^2}\left\{3\left|B_1\right|^2+12\left|B_2\right|^2+4\left|B_6\right|^2-4\mathrm{Re}\left[B_1\left(B_2^*+B_6^*\right)\right]\right\}\,,
\end{equation}
where
\begin{equation}
  \label{yeff}
  g_{eff}^2\equiv\sum Q^2(Y_s^2+Y_d^2)=\frac{52}{9}\,.
\end{equation}
The sum in (\ref{yeff}) is over all standard model charged
fermions. From ordinary running of couplings, one expects
$1/\alpha_{em}(\mathrm{1~TeV)}\sim123$ and
$1/\alpha_Y(\mathrm{1~TeV)}\sim95$. In models with extra dimensions
one generically expects power-law rather than logarithmic running
\cite{die}, which would further enhance the cross-section. However,
the deviation from the standard case only becomes significant at
energy scales considerably higher than the first KK level, so we
neglect this effect here. Fig.~\ref{sigmafig} shows the annihilation
rate as a function of the mass difference between the $\B$ and the KK
fermions.

Until now, we have only considered diagrams containing fermion
loops. There are considerably more diagrams if one also takes into
account the scalar sector of the theory; neglecting those that are
related to each other by obvious symmetries, there are actually a
total number of 22 different \emph{types} of diagrams with first KK
level modes. Using the FormCalc package \cite{formcalc}, we
implemented all these contributions numerically and found that they
merely make up a few percent of the total cross section. Comparing
this with the direct annihilation of neutralinos
$\chi\chi\rightarrow\gamma\gamma$ \cite{gg}, this is actually the
order of magnitude one would naively expect for the total cross
section when taking into account that $m_\chi/m_{\B}\sim10^{-2}$. The
large contribution from the fermion loops thus seems to be a special
feature of the KK model we have considered here. However, it is well
known that neutralino annihilation into charged leptons is helicity
suppressed and the process $\chi\chi\rightarrow\gamma\gamma$ is not
dominated by fermions, either. This indicates that the small fermion
loop contribution is a peculiarity for supersymmetry rather than its
absence a special feature of the KK model.

Higher KK levels contribute with additional diagrams, but these are
suppressed by the larger KK masses in the propagators. Numerically, we
have checked that adding second KK level fermions only changes our
previous results by a few percent. Another possibly important
contribution comes from diagrams with second KK level Higgs bosons
$H^{(2)}$ in the $s$-channel. We found, however, that for realistic
mass shifts one is too far away from the $H^{(2)}$ resonance for these
processes to be important. In conclusion, the analytical expression
(\ref{sigmav}) for the fermion box diagrams shown in
Fig.~\ref{fermions} is a good approximation for the whole annihilation
process $\B\B\rightarrow\gamma\gamma$ and we expect the effect of
other diagrams to be negligible compared to the astrophysical
uncertainties described in the next section.

One should also note that there are two other final states of $\B\B$
annihilation, $Z\gamma$ and $H\gamma$, which serve as sources of
monochromatic photons. Due to the high mass of the $\B$, these can
most likely not be discriminated from those from two-photon
annihilation. The fermionic contribution to the $\B\B\rightarrow Z\gamma$ cross section is easy to compute for a given diagram. The only difference compared to the two photon case is that the Z coupling to fermions contains both a vector \emph{and} an axial vector. From the values of these couplings we have estimated, and checked numerically, that this gives a 10\% enhancement of the gamma-ray signal coming from $\gamma\gamma$ annihilation alone. The $H \gamma$ line is left for future studies.

\section{Observational prospects}
\label{obs}

The detection rate of $\B$ annihilations in the GC depends strongly on
the details of the galactic halo profile which, unfortunately, is to a
large extent unknown. High resolution N-body simulations favor cuspy
halos, with radial density distributions ranging from $r^{-1}$
\cite{nfw} to $r^{-1.5}$ \cite{moo} in the inner regions. It has
further been suggested that adiabatic accretion of dark matter onto
the massive black hole in the center of the Milky Way may produce a
dense spike of $r^{-2.4}$ \cite{gon}. This has, however, been
contested
\cite{pierospike,milos}.  On the other hand, adiabatic contraction
from the dense stellar cluster, which is measured to exist near the
GC, is likely to compress the halo dark matter substantially
\cite{primack,klypin}. This means that there is support for a rather
dense halo profile very near the center - something that may be tested
with the new 30m-class telescopes \cite{ghez}.  Bearing these
uncertainties in mind, we will use the moderately cuspy ($r^{-1}$)
profile by Navarro, Frenk and White (NFW) \cite{nfw}.

The gamma-ray flux from WIMP annihilation in the GC is given by \cite{BUB}
\begin{equation}
  \label{phi}
  \Phi_\gamma(\Delta \Omega) \simeq
  2.92\cdot10^{-11}\, 
  \bigg( \frac{\sigma v}{10^{-29} \text{~cm}^3
  \text{\,s}^{-1}} \bigg)
  \times \left( \frac{0.8\text{~TeV}}{m_{\B}} \right)^2 \langle J_{GC}
  \rangle_{\Delta \Omega}\, \Delta \Omega \text{~m}^{-2}
  \text{\,s}^{-1},
\end{equation} 
where $\sigma v$ is the annihilation rate, and $\langle J_{GC}
\rangle_{\Delta \Omega}$ is a dimensionless line-of-sight integral
averaged over $\Delta \Omega$, the angular acceptance of the
detector. This way of writing the flux is convenient because it
separates the particle physics of the particular WIMP candidate from
the details of the galactic halo profile. For a NFW profile, one
expects $\langle J_{GC} \rangle_{\Delta \Omega} \Delta \Omega =
0.13\,b$ for $\Delta
\Omega = 10^{-5}$ \cite{ces}, which is appropriate for example for the
H.E.S.S. telescope \cite{hess}. Here, we follow \cite{bbeg} and allow
for an explicit boost factor $b$ that parametrizes deviations from a
pure NFW profile ($b=1$) and may be as high as 1000 when taking into
account the expected effects of adiabatic compression \cite{klypin}.

The direct annihilation $\B\B\rightarrow\gamma\gamma$ produces
mono-energetic photons with an energy that is to a very good
approximation given by $E_\gamma=E_{\B}\approx m_{\B}$. Due to the
velocities of the $\B$ particles, however, the emitted photons become
Doppler-shifted and the observed signal acquires a natural linewidth of
$v\sim10^{-3}$. Current ACTs have energy resolutions of about 10
to 20 percent \cite{cherenkov,hess}. Satellite-bourne
telescopes have better spectral resolution but are limited in energy
range; neither Integral ($E_\gamma\lesssim$ 10 MeV)
\cite{integral} nor GLAST ($E_\gamma\lesssim$ 300 GeV) \cite{glast}
can reach the energies of some hundred GeV to a few TeV that are
required in order to study the expected signatures of KK dark
matter. The next generation of high energy gamma-ray telescopes can be
expected to reach even better energy resolutions, also in the TeV
range, something which opens up exciting possibilities for the study
of line signals from dark matter annihilations as the one described
here \cite{gg,zg}.

\begin{figure}[t]
\psfrag{1}[][][0.9]{$1$}
\psfrag{2}[][][0.9]{$2$}
\psfrag{3}[][][0.9]{$3$}
\psfrag{4}[][][0.9]{$4$}
\psfrag{0.78}[][][0.9]{$0.78$}
\psfrag{0.79}[][][0.9]{$0.79$}
\psfrag{0.8}[][][0.9]{$0.8$}
\psfrag{0.81}[][][0.9]{$0.81$}
\psfrag{x}[t][][1]{$E_\gamma~$ [TeV]}
\psfrag{y}[b][][1]{$\mathrm{d}\Phi/\mathrm{d}E_\gamma$ $~[10^{-8}~\mathrm{m}^{-2}~\mathrm{s}^{-1}~\mathrm{TeV}^{-1}]$}
\includegraphics[width=0.7\textwidth]{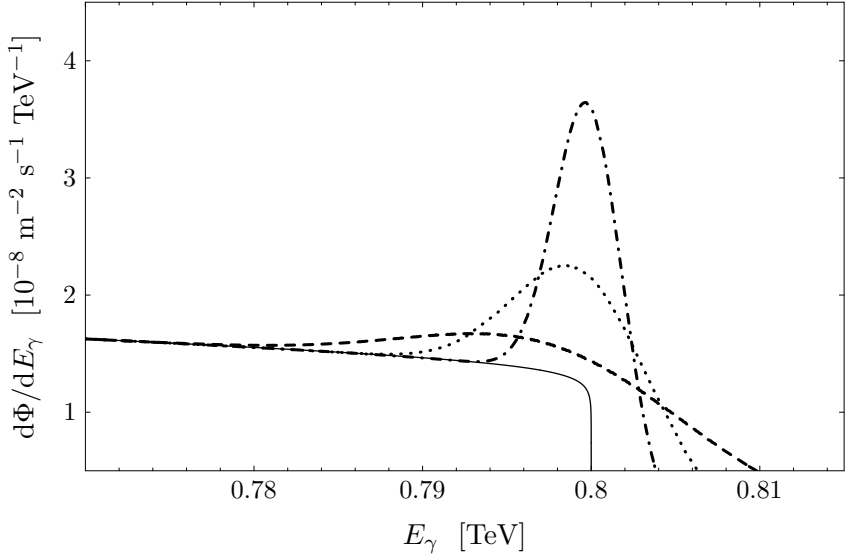}
\caption{The continuous gamma-ray flux that is expected from KK dark
matter \cite{bbeg} is plotted as a solid line. To this, we have added
the flux from direct annihilation as seen by a detector with an energy
resolution 
of $2\sigma$ = 2 \% (dashed),
1 \% (dotted) and 0.5 \% (dash-dotted), respectively. The actual
linewidth of the signal is about $10^{-3}$, with a peak value of
$1.5\cdot10^{-7}~\mathrm{m}^{-2}~\mathrm{s}^{-1}~\mathrm{TeV}^{-1}$. For
all cases, we have assumed $m_{\B}=800$ GeV, a mass shift
$m_{\xi^{(1)}}/m_{\B}=1.05$, and a moderate boost factor of $b=100$.}
\label{peakobs}
\end{figure}

Recently, the continuous gamma-ray spectrum from KK dark matter
annihilation was found to have a very promising signature \cite{bbeg},
an almost flat energy distribution that abruptly falls off at
$m_{\B}$. Including the line signal that we have studied here, one
finds a total spectrum that has an even more distinctive
signature. This is shown in Fig.~\ref{peakobs}. As one can see, the
next generation of experiments may well be able to see it -- or
falsify the KK dark matter hypothesis. Note that the
\emph{relative} strength of the line signal as compared to the
continuous signal does not depend on the astrophysical uncertainties
encoded in $b$.

In the particular model we are looking at, a dark matter particle mass
as low as 300 GeV is disfavored by the WMAP relic density
bound. However, it could be that the $\B$ does not make up \emph{all}
of the dark matter; one could also imagine a different model, with
equally non-suppressed couplings to fermions, that predicts a somewhat
higher relic density. In both cases one might be able to push the dark
matter particle mass down to a value such that an annihilation signal
like in Fig.~\ref{peakobs} could be seen already with the GLAST
satellite, taking into account its projected spectral resolution at
the highest energies.

\section{Conclusions}
\label{conc}

We have studied the possible distinct signature of annihilation into
mono-energetic gamma-rays of KK dark matter, one of the most
interesting dark matter candidates that has appeared in recent
years. We have found that with a detector of percent level energy
resolution, something that is difficult but not impossible to achieve,
such a ``smoking gun'' signal could indeed be detectable. The mass
range of the KK models that derive from UED, around 0.5 - 1 TeV,
places the new generation of ACTs at the center of attention. If the
halo density profile has a slope of $1/r$ or more and is furthermore
processed by interaction with the baryonic matter near the center of
the galaxy, then this region is the obvious place to search for the
gamma-ray line. On the other hand, it may be that the galaxy halo is
clumpy \cite{clumps} and then the best choice may be to aim at the
nearest dark matter clump. The upcoming satellite experiment GLAST
will have difficulties detecting energies above 300 GeV, but may do an
all-sky search for the continuum signal expected in these models. The
corresponding point sources without optical counterpart would be
obvious targets for the observing programs of the different ACTs then
in operation. An observation of the photon line would then
unambiguously identify the dark matter origin, as well as give the
mass of the lightest KK particle.

Much of the phenomenology would be the same for other models which do
not involve Majorana fermions as dark matter. That exception happens
on the other hand for neutralinos in supersymmetric models. As the
Majorana nature means that the annihilation into fermions is helicity
suppressed, one has to rely on a strong coupling to $W^\pm$ in the
loop to obtain large rates for the gamma line. Although there are
examples in supersymmetric models of such a coupling, for example in
the pure higgsino limit \cite{BUB} or in split supersymmetry
\cite{split}, it is not a generic feature.

In models with non-suppressed couplings to fermions, on the other hand,
the calculation of the continuous gamma-ray spectrum as given in
\cite{bbeg} would be essentially unchanged, as would the present loop 
calculation of the gamma line rate. In particular, the relative weight
of the two processes should be rather universal, meaning that whenever a
continuous signal is claimed to have been detected, there has to be a
bump at the dark matter particle mass which should stand out if the
energy resolution is a few percent, or better.

Finally, one should note that the annihilation channels
$\B\B\rightarrow Z\gamma$ and $\B\B\rightarrow H\gamma$ has a similar
structure to the one we have studied here and is another source of
monochromatic photons. We have estimated the $Z\gamma$ process to give
at least a 10\% enhancement of the signal that we have discussed in
this article.  We leave the interesting issue of the $H\gamma$ line
open for later studies.

\begin{acknowledgments}
 We are grateful to Thomas Hahn and Sabine Hossenfelder for helpful discussions.
\end{acknowledgments} 

\section*{Appendix}
\label{coefficients}

In the calculation of the loop diagrams, the following types of four-point functions appear:
\begin{eqnarray}
 D_0;D_\mu;D_{\mu\nu};D_{\mu\nu\rho};D_{\mu\nu\rho\sigma}\left(k_1,k_2,k_3;m_1,m_2,m_3,m_4\right)\qquad\qquad\qquad\qquad\qquad\qquad\qquad\nonumber\\ =
\int\frac{\mathrm{d}^nq}{i\pi^2}\frac{1;~q_\mu;~q_\mu q_\nu;~q_\mu q_\nu q_\rho;~q_\mu q_\nu q_\rho q_\sigma}{\left[ q^2-m_1^2\right]\left[(q+k_1)^2-m_2^2\right]\left[(q+k_1+k_2)^2-m_3^2\right]\left[(q+k_1+k_2+k_3)^2-m_4^2\right]}\,.
\end{eqnarray}
As is well-known, all these tensor integrals $D_{ij}$ can be reduced
to scalar loop integrals \cite{pave}, for which closed expressions
exist \cite{hoo}. In certain kinematical regimes, however, the
original reduction scheme \cite{pave} of Passarino and Veltman breaks
down. This is for example the case for the process that we are interested
in, where the two incoming $\B$ particles have essentially identical
momenta. We use the algebraic reduction scheme implemented in the LERG
program \cite{lerg} that adopts an extended Passarino-Veltman scheme
which can cope with special situations like this.

We find that all the integrals one has to calculate for the first KK
level fermion loops can be reduced to only six basic scalar
integrals. Using the conventions from \cite{hoo,lerg} (note, however,
that our signature is different from theirs), the form factors defined
in section \ref{pol} become
\begin{eqnarray}
 B_1 &=& \frac{8 \eta^2}{\eta-1}C_0(4,0,0;\eta,\eta,\eta)-\frac{8}{\eta-1}C_0(1,0,-1;0,\eta,\eta)-\frac{16}{3(\eta+1)}B_0(4;0,0)\nonumber \\
  &&-\frac{4 (\eta - 4)}{3 (\eta - 1)}B_0(4;\eta,\eta)+\frac{2(5 \eta^2 + 2 \eta - 19)}{3(\eta^2-1)}B_0(1;0,\eta)-2B_0(-1;0,\eta)+\frac{8}{3}\,,\\
 B_2 &=& \frac{4 \eta (\eta + 2)}{\eta-1}C_0(4,0,0;\eta,\eta,\eta)-\frac{4 \eta (2\eta + 1)}{\eta-1}C_0(1,0,-1;0,\eta,\eta)-\frac{8}{3(\eta+1)}B_0(4;0,0)\nonumber \\
  &&+\frac{2 (5 \eta + 4)}{3 (\eta - 1)}B_0(4;\eta,\eta)-\frac{13 \eta^2 + 10 \eta + 13}{3(\eta^2-1)}B_0(1;0,\eta)+B_0(-1;0,\eta)-\frac{8}{3}\,,\\
 B_6 &=& -\frac{4 \eta (\eta + 2)}{\eta-1}C_0(4,0,0;\eta,\eta,\eta)+\frac{4 (3\eta^2 +\eta - 1)}{\eta-1}C_0(1,0,-1;0,\eta,\eta)-\frac{8}{3(\eta+1)}B_0(4;0,0)\nonumber \\
  &&-\frac{2 (13 \eta - 4)}{3 (\eta - 1)}B_0(4;\eta,\eta)+\frac{41 \eta^2 + 26 \eta - 31}{3(\eta^2-1)}B_0(1;0,\eta)-5B_0(-1;0,\eta)+\frac{4}{3}\,,
\end{eqnarray}
where $\eta\equiv\left(\frac{m_{f^{(1)}}}{m_{\B}}\right)^2$. We
calculated the scalar loop integrals appearing in the expressions
above as follows:
\begin{eqnarray}
C_0(4,0,0;\eta,\eta,\eta) &=& -\frac{1}{2}\arctan^2\left(\frac{1}{\sqrt{\eta-1}}\right)\,,\\
C_0(1,0,-1;0,\eta,\eta) &=& \frac{1}{2}\left\{\mathrm{Li}_2\left(-\frac{1}{\eta}\right)-\mathrm{Li}_2\left(\frac{1}{\eta}\right)\right\}\,,\\
B_0(4;0,0) &=& 2 - 2 \log 2 + i \pi\,,\\
B_0(4;\eta,\eta) &=& 2-\log \eta - 2\sqrt{\eta-1}\,\arctan\left(\frac{1}{\sqrt{\eta-1}}\right)\,,\\
B_0(1;0,\eta) &=& 2- \eta \log \eta +(\eta-1) \log\,(\eta-1)\,,\\
B_0(-1;0,\eta) &=& 2+ \eta \log \eta - (\eta+1)\log\,(\eta+1)\,,
\end{eqnarray}
where the Spence function or dilogarithm is defined by
\begin{equation}
  \mathrm{Li}_2(z)\equiv-\int_0^1\mathrm{d}t\,\frac{\log(1-zt)}{t}=\sum_{k=1}^\infty \frac{z^k}{k^2}\,.
\end{equation}


\end{document}